# Write error rate of spin-transfer-torque random access memory including micromagnetic effects using rare event enhancement

Urmimala Roy[1, 2,*], Tanmoy Pramanik[1,*], Leonard F. Register[1], and Sanjay K. Banerjee[1]

[1]Microelectronics Research Center, The University of Texas at Austin, Austin, TX 78758, USA
[2]TDK-Headway Technologies, Inc., 682 S. Hillview Drive, Milpitas, CA 95035, USA
* These authors contributed equally to this work.

**Spin-transfer-torque random access memory (STT-RAM) is a promising candidate for the next-generation of random-access-memory due to improved scalability, read-write speeds and endurance. However, the write pulse duration must be long enough to ensure a low write error rate (WER), the probability that a bit will remain unswitched after the write pulse is turned off, in the presence of stochastic thermal effects. WERs on the scale of $10^{-9}$ or lower are desired. Within a macrospin approximation, WERs can be calculated analytically using the Fokker-Planck method to this point and beyond. However, dynamic micromagnetic effects within the bit can affect and lead to faster switching. Such micromagnetic effects can be addressed via numerical solution of the stochastic Landau-Lifshitz-Gilbert-Slonczewski (LLGS) equation. However, determining WERs approaching $10^{-9}$ would require well over $10^9$ such independent simulations, which is infeasible. In this work, we explore calculation of WER using "rare event enhancement" (REE), an approach that has been used for Monte Carlo simulation of other systems where rare events nevertheless remain important. Using a prototype REE approach tailored to the STT-RAM switching physics, we demonstrate reliable calculation of a WER to $10^{-9}$ with sets of only approximately $10^3$ ongoing stochastic LLGS simulations, and the apparent ability to go further.**

## I. INTRODUCTION

SPIN-TRANSFER-TORQUE RANDOM ACCESS MEMORY (STT-RAM) is a promising candidate as a "universal memory" owing to the potential for better scalability to smaller technology nodes, faster access and lower power consumption. The write process in an STT-RAM bit is inherently stochastic due to thermal fluctuations, which give rise to a distribution of the magnetization of the free layer both before and during switching. As a result, the time taken by the bit to switch has a wide distribution [1], [2]. Therefore, there will be a non-zero probability that when a finite duration write pulse is turned off the bit still will not have been written and a so-called write error will have taken place [1]. The probability that a write error takes place for a given applied current pulse of a given length is called the write error rate (WER). For correct operation of the STT-RAM array, the WER needs to be less than $10^{-9}$ if there is an error correction circuit (ECC) in the chip. If there is no ECC the WER needs to be less than $10^{-19}$ [1]. As a result, accurate modeling of the low probability tail of the WER is critical.

WER of STTRAM bits can be modeled precisely within the macrospin approximation using the Fokker-Planck (FP) method [3]. Landau-Lifshitz-Gilbert-Slonczewski (LLGS) simulations within the macrospin limit with a stochastic thermal field added also have been performed with up to ~$10^5$ independent switching trials to model switching time distributions for in-plane bits [4]. However, experimentally observed effects such as sub-volume excitations [5] and the branching of the WER plots and associated higher than otherwise expected WER [6], presumably due to higher order spin wave modes [7], cannot be captured within the macrospin approximation. For an in-depth understanding and accurate prediction of the low probability tails of WER, micromagnetic effects must be taken into account. . Previously WER, calculation including micromagnetic effects have been carried out using 64 and $10^3$ independent stochastic simulations in Ref. [8] and Ref. [9] respectively. However, the extreme tails of WERs cannot be captured in this way for micromagnetic or even macrospin simulations.

In this work, we explore calculation of WERs using stochastic LLGS with "rare event enhancement" (REE). REE long has been used for Monte Carlo simulation of other systems where rare events nevertheless remain important [10], [11]. It artificially enhances the rate of occurrence of low-probability events while proportionately reducing their weights. In Section II we describe a still prototype REE method tailored to the STT-RAM switching physics and illustrate it with macrospin calculations, which allows comparison to reference Fokker-Planck results. In Section III we provide results for full *micromagnetic* stochastic LLGS simulations, demonstrating the ability to reliably predict WER to $10^{-9}$ and likely beyond for sets of only ~$10^3$ ongoing LLGS simulations, with switching currents consistent with practical usage [2], [12]. Nevertheless, we emphasize that this prototype represents a first, proof-of-concept approach; we expect that there remains significant room for improvement.

## II. RARE EVENT ENHANCEMENT FOR WERs AND COMPARISON TO FOKKER-PLANK RESULTS WITHIN THE MACROSPIN LIMIT

### A. Basic Method

We employ a REE method known as "Importance splitting" [13]–[16]. The core idea is that for many stochastic systems, before the system reaches some state of extremely low level of probability $L_N$, it traverses multiple intermediate higher levels of probability with more limited relative separation between each subsequent level [16]. Thus, extremely rare events can be reached with high probability by from time to time splitting "parent" trajectories that are more likely to lead to rare events into multiple "offspring" trajectories, but each of corresponding lower weight than the parents to conserve the norm. Each offspring trajectory then follows its own stochastically independent trajectory. Beyond this point, there is both much generality and detail to importance splitting methods and their optimization, which is beyond the scope of this work. Our goal here is only to demonstrate a specific, relatively simple, prototype adaptation to stochastic LLGS simulation of STT-RAM switching as a proof-of-concept and starting point.

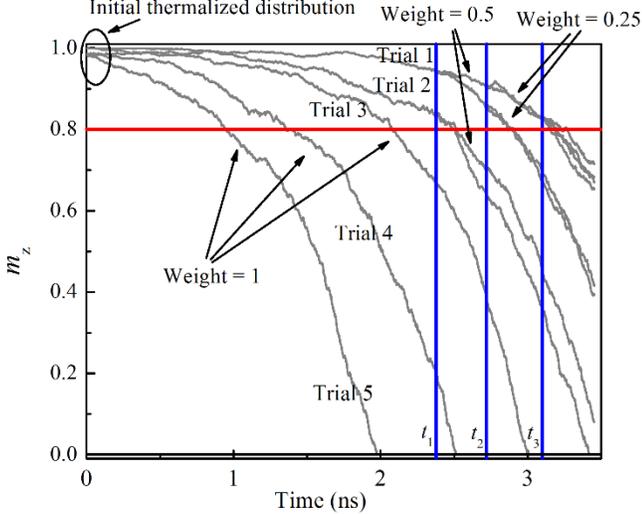

FIG. 1. Enhancement of rare stochastic LLGS magnetization trajectories of the (normalized) $m_z$ (divided by the magnitude of the total magnetization of the bit) at discrete times $= t_1, t_2, t_3 \ldots$ . (These specific trajectories are for a bit of circular cross-section with perpendicular magnetic anisotropy (PMA) in the macrospin approximation, but at least for this prototype proof-of-concept approach our basic method is the same for micromagnetic simulations.) STT switching is initiated at $t = 0$ ns from a thermalized distribution with $m_z$ close to unity. Starting with $m_z$ near but not at 1 due to the stochastic thermal field, the bit magnetization trajectories are pushed toward $m_z = -1$ due to STT but still subject to the stochastic thermal field, as the time progresses. At each enhancement time (vertical lines, blue online), "parent" trajectories with $m_z$ above a predetermined threshold value of $m_z$ (horizontal line, red online) are split into two "offspring" of weight equal to half the weight of the parent trajectories; trajectories with $m_z$ below the threshold but above zero are continued as they are (1 offspring); and trajectories that have reached $m_z = 0$ or below are not continued (no offspring).

The first critical order of business is to choose a predictor of which trajectories are more likely to lead to rare events than others. For this STT-RAM, the state of the bit can be characterized reasonably by the surface-normal component of the magnetization normalized to the magnitude of the total magnetization, $m_z$ such that $1 \geq m_z \geq -1$. At any time $t$, the greater $m_z(t)$, the smaller the in-plane component of the (normalized) magnetization $\mathbf{m}(t)$, the smaller the STT on the bit, and the slower the bit is being accelerated toward $m_z = -1$, and, thus, the longer switching is likely to take from that time forward. Moreover, the closer $m_z(t)$ is to unity, the greater the relative importance of the continuing stochastic thermal field as compared to the STT, and, thus, the less deterministic at least near future trajectories $\mathbf{m}(t)$ will be, where splitting completely deterministic trajectories serves no purpose even if that trajectory is likely to lead to a rare event. Thus, we chose $m_z$ as our measure of the likelihood that a magnetization trajectory $\mathbf{m}(t)$ will ultimately lead to a rare non-switching event, as well as the potential usefulness of importance splitting of the trajectory.

The prototype REE procedure used here is illustrated in Figure 1. This figure and the chosen trajectories are optimized for conceptual clarity and not results. These trajectories also are for a bit of circular cross-section with perpendicular magnetic anisotropy (PMA) in the macrospin approximation, but for this prototype method the approach is the same for micromagnetic simulations. The simulation of switching is subdivided into time intervals of variable length $\tau_l = t_{l+1} - t_l$. These time intervals are adjusted on the fly such that, for a substantial set of trajectories $\mathbf{m}(t)$ considered simultaneously, the time interval is terminated when the number of trajectories for which $m_z$ has fallen below zero at the end of the time interval is approximately equal the number of trial for which $m_z$ is closer to unity than a chosen $m_z$ threshold value $m_{z,\text{th}}$ (the choice of which will be returned to below). Then, at the beginning of each new time interval beyond the first, the "parent" trajectories $\mathbf{m}(t)$ with $m_z > m_{z,\text{th}}$ are subdivided into two "offspring" trajectories, each weighted by one-half the weight of the parent trajectory, which then continue along their own stochastically independent subsequent trajectories; trajectories with $m_{z,\text{th}} > m_z > 0$ are continued as they are with the same weighting; and trajectories with $m_z < 0$ are considered switched and are discontinued. (Trajectories with $m_z < 0$ are considered beyond the point of no return *if* the pulse were to continue. That some trajectories with $m_z < 0$ might not end up switched if the write pulse were turned off at this point or, conversely, that some trajectories with $m_z > 0$ would nevertheless end up switched if the write pulse were turned off at this point should not noticeably effect the WER within the resolution of interest here.) In principle, the total number of ongoing trajectories would be conserved precisely at all times in this way if we could terminate time intervals when the number of trajectories with $m_z < 0$ and the number with $m_z > m_{th}$ were precisely equal. In practice, however, we have to stop and restart all of the stochastic LLGS simulations to check these numbers, which so far we have done at fixed small but nonzero time intervals. As a result we have allowed for a limited inequality, and the total number of ongoing trajectories can differ somewhat from the original through time. However, we also adjust the inequality window about the equality through time to bias the number of ongoing trajectories back toward the original value, so that the number of trajectories $\mathbf{m}(t)$ and—the associated computational effort solving the stochastic LLGS simulations—remains effectively conserved through simulation time. However, the enhancement of rare trajectories at the expense of more common trajectories during each time interval leads to an ultimately strong artificial skewing of the magnetization trajectories towards the rare events of interest, with the possibility of a given individual trajectory remaining into the $l^{\text{th}}$ time interval $\tau_l$ having been enhanced by up to $2^l$ in the limit.

For Fig. 1, specifically, a small illustrative sample of magnetization dynamics trajectories has been plotted vs time along with a relatively low (far from unity) threshold $m_{z,th} = 0.8$ used for visual clarity. Trial trajectories 1 and 2 have $m_z(t) > m_{z,th}$ at the beginning of the first enhancement $t_1$—a time determined considering many more trajectories not shown—and, hence, each of them are split into 2 offspring trajectories, each with their weights reduced from unity to 0.5. Trial trajectories 3 and 4 have $0 < m_z < m_{z,th}$, and hence advance as they are without any enhancement. Trial trajectory 5 has reached a negative value of $m_z$ at $t_1$ (not shown) and is discontinued. At time $t_2$ both offspring trajectories of Trial trajectory 1 are again split in two offspring trajectories each, and each with a weight now reduced to 0.25, while Trial trajectory 4 is discontinued. This procedure continues at $t_3$ and beyond.

Like the choice of the predictor of low probability events, the choice of the threshold value of the predictor also is important. It also is non-trivial. To this end we now turn to a study of WER for macrospins calculated with our REE method. These macrospin simulations are, of course, computationally less demanding than micromagnetic simulations and, thus, allows for more simulations and larger data sets. They also allow for comparison to reference Fokker-Planck results.

### B. Choosing $m_{z,th}$ and illustration of REE of macrospin simulations

For all the simulations in this work, macrospin or micromagnetic, a thermal fluctuation field [17] was added to the effective field term in the LLGS equation. The scheme of Heun [18] was used for integrating the stochastic LLGS equation with an integration time step of 1 fs for these macrospin simulations. The simulation temperature was set to $T = 300$ K. Initial ($t = 0$) equilibrium thermal distributions of **m**($t$) were obtained by starting with a value of $m_z = 0$ at $t = -5$ ns and performing the stochastic LLGS simulation up to $t = 0$ ns with no STT and no REE.

For the macrospin bit free layer, a saturation magnetization $M_s = 1.1 \times 10^6$ A/m, a uniaxial anisotropy energy $K_u = 8 \times 10^5$ J/m$^3$ with axis along the perpendicular direction, and a Gilbert damping constant $\alpha = 0.01$ were used. The diameter and thickness of the bit were taken to be 56 nm and 1 nm, respectively. Assuming a rectangular prism [19], the demagnetization factors are found to be $N_{xx} = N_{yy} = 0.027$ and $N_{zz} = 0.946$. The thermal stability factor $\Delta$ of the PMA bit is given by $\Delta = \frac{V}{k_B T}\left[K_u - \frac{1}{2}\mu_0 M_s^2 (N_{zz} - N_{xx})\right]$, where $V$ is the volume of the bit, $k_B$ is the Boltzmann's constant, $\mu_0$ is the vacuum permeability. With the above material parameter values and dimensions, $\Delta$ of the simulated PMA bit becomes 60. The fixed layer was considered only as a source of spin-polarized current that flows perpendicular to the plane of the free layer. (Essentially, we assumed use of synthetic antiferromagnet stack for the fixed layer to minimize the dipolar interactions between the fixed and free layer). The critical switching current $I_{c0}$ for STT switching can be written in terms of $\Delta$ as, $I_{c0} = (4\alpha q \Delta k_B T)/\eta \hbar$, with $\eta$ being the spin torque efficiency factor (spin polarization factor) from the fixed layer. Taking $\eta = 0.4$ gives $I_{c0} = 37.7$ μA.

Figure 2 shows the simulated probabilities as a function of time of not having switched $P_{ns}$—essentially the write error rate if switching current were turned off at that time as discussed in Subsection A—for the PMA bit obtained with REE for three different values of $m_{z,th}$ as shown, as well as without REE for reference. The switching current is $2I_{c0}$. Each of the, here, 100 simulation sets consist of $10^3$ independent trials. Here a "trial" corresponds to a single initial trajectory, which subsequently may or may not give rise to one or more offspring trajectories. As the total weighting including continuing and switched (terminated) trajectories is conserved for each trial, the total weighting for all trajectories springing from all trials in a set is $10^3$ at all times. Therefore, $P_{ns}$ for each set of trials can be written,

$$P_{ns} = \frac{\sum_{not-switched\ trajectories} Weight}{\sum_{all\ trajectories} Weight}$$
$$= \frac{\sum_{not-switched\ trajectories} Weight}{10^3}$$

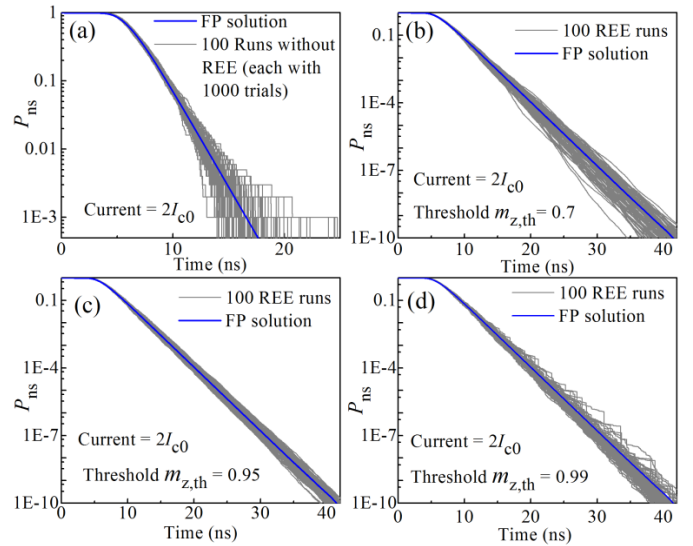

FIG. 2 Effect of selection of threshold $m_{z,th}$ used for REE on the calculated WER for a macrospin with an applied current of twice the critical current, $I_{c0}$, (thin lines, grey online) with the Fokker-Planck solution provided for reference (thick lines, blue online). The results of each figure represent 100 sets (each represented by its own grey line) of $10^3$ independent trials. (a) Results without REE for reference. Intrinsically, the lowest non-zero value of WER for each set is $10^{-3}$ as seen. (b), (c) and (d) Results with REE with thresholds, $m_{z,th} = 0.7$, 0.95 and 0.99, respectively. All choices of $m_{z,th}$ provide vastly improved estimations of WER as compared to simulations without REE, but $m_{z,th} = 0.95$ clearly provides the best of the three.

As can be seen in Fig. 2, use of REE with any of the choices of $m_{z,th}$ provides vastly improved estimations of WER. However $m_{z,th} = 0.95$ clearly provides the least variability among these choices. Consistent with the discussion of Subsection A, if $m_{z,th}$ is too low (further from unity), then, we waste REEs on too many trajectories that are already largely deterministic. If $m_{z,th}$ is too high (closer to unity), we miss REEs for too many trajectories that remain largely stochastic. Indicative of this latter limit is the quasi-step like behavior for many of the trial sets, noticeable at fairly large WERs but increasing in scale as WER decreases, as seen in Fig. 2(d), partially reminiscent of the behavior for the REE free result of Fig. 2(a) if at much smaller WERs.

Toward identifying the optimum threshold a priori, a significant sample of individual trials—$10^3$ again for our simulations—of $\mathbf{m}(t)$ were simulated *without* REE. For each of these trials a "transit" time was identified as a function of $m_{z,th}$, which is defined as the time required for $\mathbf{m}(t)$ to fall from $m_z = m_{z,th}$ to $m_z = 0$ (and, similarly, an "incubation" time is identified as the time for $\mathbf{m}(t)$ to reach $m_{z,th}$ from $t = 0$) consistent with prior simulation work [20]–[22]. We then considered the standard deviation in the transit time among all trials. We found that the inflection point in this curve, as shown in Fig. 3, provides a reasonably optimal value of $m_{z,th}$ as the inflection signifies a transition between more stochastic paths with thus large variability in transit times, and more deterministic paths with thus similar transit times. In particular, the 0.95 value for $2I_{c0}$ that was predicted in this ways is consistent with the results of Fig. 2. Moreover, as can be seen, the optimum value of $m_{z,th}$ increased toward unity with increased applied current. The greater the current, the greater the STT for a given value of $m_z$, the less deviation of $m_z$ from unity that is required for the STT to overcome the thermal fluctuation field on average.

Figure 4 shows the WER for 100 sets of $10^3$ trials each obtained with REE using the optimum thresholds found from Figure 3 for applied currents of $I = 2I_{c0}$, $3I_{c0}$ and $4I_{c0}$, respectively, along with error bars representing a plus or minus two standard deviations $\pm 2\sigma$ (95%) confidence interval for the *individual* trial sets based on the variation among them. The $\pm 2\sigma$ confidence interval *for the average result of all 100 sets of trials* is an order of magnitude smaller still. This later average of the predicted $P_{ns}$ from the all REE enhanced stochastic LLGS trials follows the exact Fokker-Planck solution quite well.

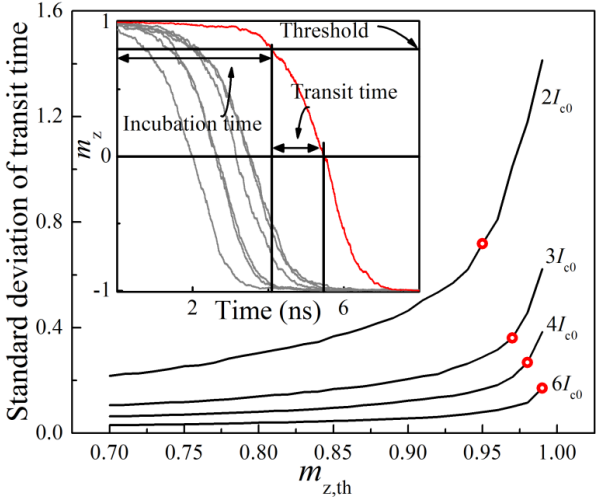

FIG. 3: Variation of transit time as a function of the threshold. The definition of the transit time for a given threshold $m_{z,th}$ is illustrated in the inset for one trial trajectory (highlighted in red online). The optimum threshold for each current is marked as a (red online) circle on the corresponding curves.

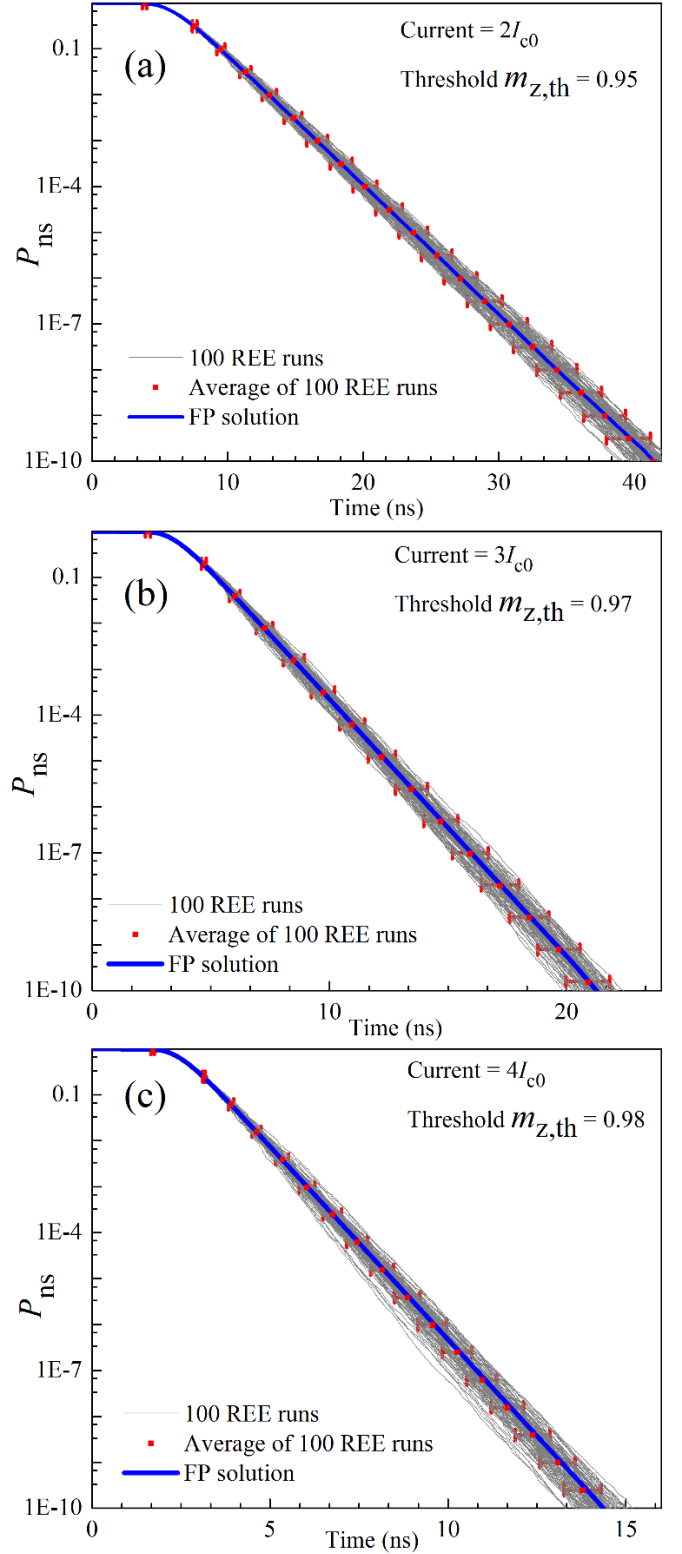

FIG. 4: REE simulation of WER using a current dependent optimized $m_{z,th}$ threshold for a switching current of (a) $2I_{c0}$ with $m_{z,th} = 0.95$, (b) $3I_{c0}$ with $m_{z,th} = 0.97$ (c) $4I_{c0}$ with $m_{z,th} = 0.98$. Error bars (red online) represent plus or minus two standard deviations among the samples. Fokker-Planck results are provided for comparison.

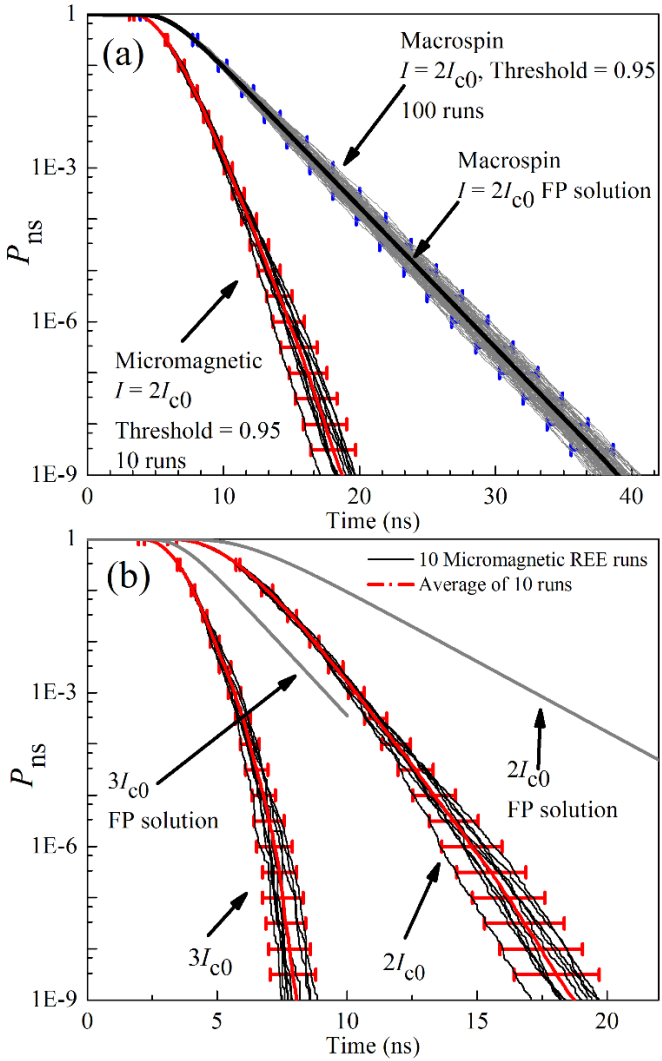

FIG. 5. WER obtained from micromagnetic simulations with REE. (a) WER calculated using REE for $I = 2I_{c0}$ using a threshold of 0.95 for both micromagnetic and macrospin calculations. For the micromagnetic WER calculation, 10 independent REE runs are shown along with the average of 10 such runs. (b) 10 independent REE runs with micromagnetic simulations for $I = 2I_{c0}$ ($m_{z,\text{th}} = 0.95$) again and $I = 3I_{c0}$ ($m_{z,\text{th}} = 0.95$). Fokker-Plank solutions for the same bit are also plotted. (The Fokker-Planck solution for $I = 3I_{c0}$ is not shown after 10 ns for visual clarity.) The horizontal bars in (a) and (b) to the left and right of the WER data points indicate plus or minus two times expected standard deviation among the sets of $10^3$ trials in the time taken to reach the corresponding WER, calculated using 10 runs (100 runs) for the micromagnetic (macrospin) calculations.

## III. MICROMAGNETIC STOCHASTIC LLGS WITH RARE EVENT ENHANCEMENT

### A. Method

The GPU-based micromagnetic simulator MUMAX3 [23] was used to carry out the micromagnetic simulation of STT switching of a PMA bit. The diameter and thickness of a circular free layer were taken to be to be 60 nm and 1 nm respectively, just slightly larger in diameter than that for the preceding macrospin simulations. The grid size was taken to be 1.88 nm in the plane of the free layer and 1 nm perpendicular to the plane of the free layer. Grids were coupled through an exchange constant $A_{\text{ex}}$ of 20 pJ/m. (An infinite $A_{\text{ex}}$ represents a macrospin.) Otherwise, the parameters of the free magnet were taken as the same as for the macrospin simulations. With these parameters, the thermal stability factor $\Delta$ of the bit is ~67 in the limit of coherent rotation as for the macrospin. The corresponding macrospin critical switching current $I_{c0}$ calculated as before is 42.1 µA. These calculated values of thermal stability and the critical current may well be overestimates of the true values for the micromagnetic system, but still can be used as a reference. The temperature again was set to be 300 K. A 10 fs integration step size was used for all the micromagnetic simulations. As for the macrospin trials, the initial population of trial trajectories was thermalized by simulating magnetization dynamics for 5 ns under the influence of thermal fluctuations before the switching current was applied at $t = 0$. For these micromagnetic REE simulations, as a simple extension of the approach used for the macrospin simulations, the predictor of which trajectories are more likely to lead to rare events than others was taken as $\langle m_z(x, y, t)\rangle$, the spatial average of $m_z(x, y, t)$ normalized to the magnitude of $\langle \mathbf{m}(x, y, t)\rangle$. Then, also as for the macrospin simulations, a threshold value of $\langle m_z(x, y, t)\rangle$, $m_{z,\text{th}}$, was used to identify the rare, slow switching trials.

### B. Results

Figure 5(a) shows the WER calculated for $I = 2I_{c0}$ for $m_{z,\text{th}} = 0.95$, which again represents the pre-calculated inflection point in the standard deviation of the previously defined transit time. Each of the 10 individual lines represents a set of $10^3$ independent REE simulation trials. Also plotted in Figure 5(a) is the predicted WER from 100 sets of $10^3$ REE macrospin trials the same system otherwise, with accuracy again verified by comparison to the Fokker-Planck result. First, of course, we note that a macrospin analysis quantitatively overestimates the required switching time for a given switching current substantially. In terms of the subject at hand, we note that WERs can be predicted using REE down to $10^{-9}$ and likely beyond with sets of only $10^3$ independent REE trials, with good reliability as evidenced by the $\pm 2\sigma$ (95%) confidence interval for the individual trial sets based on the variation among them. Again, the $\pm 2\sigma$ confidence interval for the average result of all sets of trials is smaller, by a factor of 3 here. In absolute terms, the $\pm 2\sigma$ confidence interval in time to achieve a given WER for the macrospin results is much the same as for the micromagnetic results, although the error in the micromagnetic simulations relative to the average is significantly larger. Also the REE runs with a threshold of 0.95 show some step-like artifacts similar to the ones shown in Figure 2(d), suggesting that the employed value of $m_{z,\text{th}} = 0.95$ may be a bit too close to unity. However, lowering $m_{z,\text{th}}$ to 0.9 creates a larger variation (not shown) between the predicted results, qualitatively similar to the results of Fig 2(b) as compared to those of Fig. 2(c).

The results of the REE micromagnetic simulations with $I = 3I_{c0}$ and, for reference, $I = 2I_{c0}$ again are shown in Fig. 5(b). The Fokker-Plank results for each value of switching current $I$ also are shown for reference. The value of $m_{z,\text{th}}$ used for the shown $I = 3I_{c0}$ results is 0.95 which seems to be near optimal, already showing a bit of the step like behavior, while an $m_{z,\text{th}} = 0.97$ value is suggested by the standard deviation of the transit time. Again lowering $m_{z,\text{tb}}$ to 0.9 produces more variation. Moreover, it becomes increasing evident that, as the switching current increases, the $P_{ns}$ decay becomes super-exponential (super-linear on these log-linear plots) qualitatively deviating from macrospin behavior.

For practical applications $3I_{c0}$ already may be considered a large switching current [2], [12], particularly given that the employed $I_{c0}$ obtained in the macrospin limit may well be a significant overestimate of the true $I_{c0}$ as previously noted. Nevertheless, it is informative from a modeling perspective to consider what happens as we follow the switching current higher (not shown). As currently simulated, for $I = 4I_{c0}$, we were unable to reliably predict WER below ~$10^{-6}$, and for $I = 6I_{c0}$ below ~ $10^{-5}$. Some of the trial sets, each with initially $10^3$ samples, became null, i.e. none of the trajectories remained unswitched. With the increasingly quick reduction in the WERs with increasing current and simulated time, combined with nonzero periods for sampling the subsets of trajectories with $m_z > m_{z,th}$, $m_{z,th} > m_z > 0$ and $0 > m_z$, it became increasingly difficult maintain fixed trial set size. Thus, increasing the resolution in the time for checking these subset of trajectories, by brute force or refined technique, should improve results in the high-current limit substantially. Moreover, it may also be that that our predictor of which trajectories are more likely to lead to rare events than others, the normalized average value of $m_z$, $\langle m_z(x,y,t)\rangle$, becomes less efficient at higher current densities. Indeed, when going from the macrospin to micromagnetic simulations we noticed some softening of the inflection in the standard deviation of the transit times by which $m_{z,th}$ is chosen (not shown), consistent with the greater relative error for the micromagnetic simulations. That result should perhaps not be surprising given that we already were mapping a three dimensional quantity, $\mathbf{m}(t)$, to one dimensional predictor, $m_z$, for the macrospin simulation. For the micromagnetic simulation with $N$ semi-independent magnetic cells, we now are mapping a $3N$ dimensional quantity, $\mathbf{m}(x,y,t)$, to a still only one dimensional predictor, $\langle m_z(x,y,t)\rangle$. It might be possible to obtain still more reliable results by more carefully considering, e.g., the quasi-independent contributions of perhaps the lowest few "odd" eigenmodes ($l$) of the magnetization $\mathbf{m}_l(x,y)$ (those for which the average magnetization does not vanish). Also, as previously noted, the REE method used here is basic; use of more sophisticated REE methods may also lead to still more reliable prediction for a given computation effort. In other words, even with the already great advantage of REE for simulation of WERs for practical micromagnetic systems demonstrated with the simple approach used here, there remains room to further improve the REE-based calculation of WER in the future.

IV. CONCLUSION

We have developed and demonstrated a relatively simple, prototype REE method tailored to calculation of WER in micromagnetic STT-RAM systems, and have demonstrated statistically reliable prediction of WER down to ~$10^{-9}$ with sample sets of only $10^3$ stochastic LLGS-based switching trials. Moreover, the total number of ongoing stochastic LLGS simulations, the computational burden, is effectively conserved through simulation time. In addition, nothing in our simulated results suggests that the use of REE in micromagnetic simulation cannot allow still lower WERs to be calculated for common switching currents, even with the basic method described. Moreover, using more refined REE methods and development of improved predictors of rare magnetization trajectories offers the opportunity for still more reliable prediction of WERs, to lower WERs, and under more extreme switching conditions.

ACKNOWLEDGMENT

U. R. thanks David L. Kencke, Angik Sarkar and Charles C. Kuo for the useful discussions on WER during her internship at Intel Corp. in the summer of 2014. The authors acknowledge the Texas Advanced Computing Center (TACC) at The University of Texas at Austin for providing high performance computing resources that have contributed substantially to the research results reported within this paper. This work was supported in part by the NRI SWAN and the NSF NASCENT ERC.